\documentstyle[epsfig]{mn}
\def\figdir{.}

\def\epsscale#1{\epsfxsize=#1\columnwidth}
\def\plotone#1{\par\centerline{\epsfbox{#1}}}

\def\etal{{ et al.~\/}}

\title{Morphology and Evolution in Galaxy Clusters I:  Simulated Clusters 
in the Adiabatic limit and with Radiative Cooling}

\author{Nurur Rahman$^1$, Sergei F.\ Shandarin$^1$, 
Patrick M. Motl$^2$, and Adrian L. Melott$^1$ \\
$^1$Department of Physics and Astronomy,
University of Kansas, Lawrence, KS 66045, USA;\\
$^2$Center for Astrophysics and Space Astronomy, 
University of Colorado, Boulder, CO 80309, USA;\\
nurur@kusmos.phsx.ukans.edu, sergei@ku.edu, 
motl@casa.colorado.edu, melott@kusmos.phsx.ukans.edu}
\date{}
  
\begin{document}
\maketitle


\begin{abstract}
We have studied morphological evolution in clusters simulated in the 
adiabatic limit and with radiative cooling. Cluster morphology in the 
redshift range, $0 < z < 0.5$, is quantified by  multiplicity and 
ellipticity. 
In terms of ellipticity, our result indicates slow evolution in 
cluster shapes compared to those observed in the X-ray and optical 
wavelengths. The result is consistent with Floor, Melott \& Motl 
(2003). In terms of multiplicity, however, the result indicate 
relatively stronger evolution (compared to ellipticity but still 
weaker than observation) in the structure of simulated clusters 
suggesting that for comparative studies of simulation and observation, 
sub-structure measures are more sensitive than the shape measures. 
We highlight a few possibilities responsible for the discrepancy in 
the shape evolution of simulated and real clusters. 
\end{abstract}
\section{Introduction}
The hierarchical clustering is the most popular model for the Large 
Scale Structure (LSS) formation. The model relys on the assumption 
that the larger clumps of matter distributions result due to the 
merging of smaller sub-clumps in cosmological time. Structural 
evolution in cosmological systems such as galaxies and clusters of 
galaxies is therefore the underlying principle in this scenario. 
 

Melott, Chambers \& Miller (2001; hereafter MCM) has reported evolution 
in the gross morphology of galaxy-clusters (quantified by ellipticity) 
for a variety of optical and X-ray samples over the redshift, $z < 0.1$. 
They infer that the evidence is consistent with low matter density 
universe. Using ellipticity as well as intracluster medium temperature 
and X-ray luminosity, Plionis (2002) has presented evidence for the 
recent evolution in optical and X-ray cluster of galaxies for redshift, 
$z \leq 0.18$. In both studies evolution is quantified by the change of 
cluster ellipticity with redshift. In a recent study, Jeltema et al. 
(2003) have reported structural evolution of clusters with redshift 
where cluster morphology is quantified by power ratio method (Buote 
\& Tsai 1995). The authors used a sample of 40 X-ray clusters over the 
redshift range $0.1 - 0.8$ obtained from Chandra Observatory. The results 
of these previous studies, although each one employed different techniques 
in their analyses, indicate evolution in the morphology of the largest 
gravitationally bound systems over a wide range of look-back time. 

The observational evidences prompted concerns on the formation and 
evolution of structures in numerical simulations. If the results of 
simulations give us faithful representations of the evolutionary 
history of cosmological objects than one would expect a similar trend 
in the structure of simulated objects. So far almost all studies of 
simulated clusters are focused on understanding the nature of the 
background cosmology within which the present universe is evolving 
(Jing et al. 1995; Crone, Evrard \& Richstone 1996; Buote \& Xu 1997; 
Valdarnini, Ghizzardi \& Bonometto 1999). 
Until recently a comparative study of morphological evolution in the 
simulated and real clusters was absent. Floor et al. (2003) and Floor, 
Melott \& Motl (2003; hereafter FMM) have investigated evolution in 
clusters morphology simulated with different initial conditions, 
background cosmology, and different physics (e. g., simulation with 
and without radiative cooling). They have used eccentricity as a 
probe to quantify evolution. Their studies, emphasizing on measuring 
shapes in the outer regions of clusters, suggest slow evolution in 
simulated cluster shapes compared to the observed clusters.       

In this paper we study evolution in simulated clusters in flat cold dark 
matter universe ($\Lambda$CDM; $\Omega_m = 0.3$, $\Omega_{\lambda} = 0.7$) 
by using high resolution simulations (Motl et al. 2003). 
We use two data sets: the first set have clusters simulated in the adiabatic 
limit and the other set contains clusters simulated with radiative cooling. 
Each sample contain X-ray and dark matter distributions and has been 
analyzed at four different redshifts, $z = 0.0, 0.10, 0.25,$ and $0.50$. 

This is the first in a series of papers aimed to study the morphology 
and evolution of clusters of galaxies. The papers will use shape measure 
such as ellipticity derived from the Minkowski functionals (Minkowski 1903; 
hereafter MFs). The MFs provide a non-parametric description of the images 
implying that no prior assumptions are made on the shapes of the images.     
The measurements based on the MFs appear to be robust and numerically 
efficient when applied to various cosmological studies, e. g., galaxies, 
galaxy-clusters, CMB maps etc. (Mecke, Buchert \& Wagner 1994; 
Schmalzing et al. 1999; Beisbart 2000; 
Beisbart, Buchert \& Wagner 2001; Beisbart, Valdarnini \& Buchert 2001; 
Kerscher et al. 2001a, 2001b; 
Sheth et al. 2003; Shandarin, Sheth \& Sahni 2003). Various measures, 
derived from the two-dimensional scalar, vector and several tensor MFs to 
quantify shapes of galaxy images, had been described and tested  in Rahman 
\& Shandarin (2003a, 2003b; hereafter RS1 and RS2). The multiplicity, 
however, is not constructed from the MFs (see section $\S 3$ for details 
of parameter construction).

In this paper we use the extended version of the numerical code used in 
RS1 and RS2. In the following paper in this series we will analyze 
simulated clusters with cooling and heating mechanism such as star 
formation and star formation with supernovae feedback. We will also make 
a comparative study of observed and simulated clusters with up to date 
samples of optical and X-ray clusters.

The organization of the paper is as follows: simulation technique 
is described in $\S2$, a brief discussion of shape measures is given in 
$\S3$. The results are presented in $\S4$ and the conclusions are 
summarized in $\S5$.
\section{Numerical Simulations}
We have analyzed projected clusters (in all three axes) simulated 
in the standard, flat cold dark matter universe ($\Lambda$CDM) with 
the following parameters: $\Omega_b = 0.026$, $\Omega_m = 0.3$, 
$\Omega_{\lambda} = 0.7$, $h = 0.7$, and $\sigma_8 = 0.928$ (Motl et 
al. 2003). We have used two samples of clusters derived from the same 
initial conditions and background cosmology. The major difference 
between the samples is in energy lose mechanism experience by the 
baryonic fluids. In one sample, the fluid is allowed to lose energy 
via radiation and subsequently cool; in the other sample no energy 
lose is allowed. 

The simulations use a coupled N-body Eulerian hydrodynamics code 
(Norman \& Bryan 1999; Bryan, Abel \& Norman 2000) where the dark 
matter particles are evolved by the adaptive particle-mesh, N-body 
code.  The PPM scheme (Colella \& Woodward 1984) is used to treat 
the fluid component on a comoving grid. An adaptive mesh refinement 
(AMR) is employed to concentrate the numerical resolution on the 
collapsed structures that form naturally in cosmological simulations. 
The dark matter particles exist on the coarsest three grids; each 
sub-grid having twice the spatial resolution in each dimension and 
eight times the mass resolution relative to its parent grid. At the 
finest level, each particle has a mass of 
$9 \times 10^{9} \; \mathrm{h}^{-1} \; \mathrm{M_\odot}$.
A second order accurate TSC interpolation is used for the adaptive 
particle mesh algorithm.  Up to seven levels of refinement are 
utilized for the fluid component, yielding a peak resolution of 
$15.6 \: \mathrm{h}^{-1}$ kpc within the simulation box with sides 
of length 256 $\mathrm{h}^{-1}$ Mpc at the present epoch.  

Since fluid is allowed to radiatively cool, a tabulated cooling 
curve (Westbury \& Henriksen 1992) for a plasma of fixed, 0.3 solar 
abundance has been used to determine the energy loss to radiation.  
The cooling curve falls rapidly for temperatures below 
$10^{5} \; \mathrm{K}$ and is truncated at a minimum temperature of 
$10^{4} \; \mathrm{K}$. Heat transport by conduction is neglected in 
the present simulation since it has been shown that even a weak, 
ordered magnetic field can reduce conduction by two to three orders 
of magnitude from the Spitzer value (Chandran \& Cowley 1998). 
However, Narayan \& Medvedev (2001) has shown that if the chaotic 
magnetic field fluctuations extends over a sufficiently large length 
scales within the intra-cluster medium (ICM), then thermal 
conductivity becomes significant to the global energy balance of the 
ICM. Energy input into the fluid from supernovae feedback or discrete 
sources such as AGN are also neglected in the current simulations. 
For a complete description of simulation in the adiabatic limit and 
with radiative cooling see Motl et al. (2003).  

The adiabatic and radiative cooling samples, respectively, have 43 
and 41 clusters in three dimension. Therefore we have a total of 129 
($N_{ad}$) and 123 ($N_{rc}$) projected clusters in the respective 
samples. Each projected cluster image is within a 8 h$^{-1}$ Mpc 
(comoving) frame containing 360 $\times$ 360 pixels.
\section{Morphological Parameters}
We use multiplicity ($M$) and ellipticity ($\epsilon$) as quantitative 
measures to study evolution in simulated clusters. Ellipticity is  
derived from the area tensor, a member of the MFs. The details on the 
MFs can be found in Schmalzing (1999), Beisbart (2000), and RS1.
Here we discuss briefly the construction of the measures. 

$\bullet$ Multiplicity ($M$):
This parameter is defined as, 
\begin{equation}
M = \frac{1}{A_{max}} \ \sum_{i=1}^N A_i,
\end{equation}
where $A_i$ is area of the individual components at a given level and 
$A_{max}$ is the area of the largest component at that level, and $N$ 
is the total number of components. It is a measure with 
fractional value and gives the number of components measured at any 
brightness level: $M=1$ for a single iso-intensity contour (i. e. 
component) and $M > 1$ for multi-contours. Multiplicity is different 
than the Euler Characteristic ($\chi$), a member of the scalar MFs 
(see RS1 for more). The EC is an integer number that gives the total 
number of components present at any given level but does not provide 
any information regardless of their sizes. Multiplicity, on the other 
hand, is able to extract this information.
 
It may be mentioned here that Thomas et al. (1998) have also used 
multiplicity as a parameter for sub-structure measure in N-body 
simulations. They define it as a ratio of mass of sub-clumps to cluster 
mass. In this study it is a ratio of the areas as defined in equation 1. 

We use two variants of $M$ to present our results: one is the average of 
multiplicity over all density/brightness levels, $\bar{M}_{eff}$, and 
the other is the maximum of the multiplicity found at one of the levels, 
$M_{max}$. 

$\bullet$ Ellipticity ($\epsilon$): We adopt the definition of ellipticity, 
\begin{equation} 
\epsilon = 1 - b/a,
\end{equation} 
where $a$ and $b$ are the semi-axes of an ellipse. For our purpose the 
semi-axes correspond to the ``auxiliary ellipse'' constructed from the 
eigenvalues of the area tensor (see RS1 for detail). 
We have used two variations of $\epsilon$: one is sensitive to the shape 
of the individual cluster components present at a given level while the 
other is sensitive to the collective shape formed by all the components 
present at that level. 
We label these two variants of $\epsilon$, respectively, as the effective 
(${\epsilon}_{eff}$) and the aggregate (${\epsilon}_{agg}$) ellipticity. 
Morphological properties of clusters such as shape and the nature or the 
degree of irregularity existing in these systems can be probed effectively 
with these two parameters.

At any given density/brightness level, we construct ${\epsilon}_{eff}$ 
as,
\begin{equation} 
{\epsilon}_{eff} = \frac{1}{M \cdot \ A_{max}} \ \sum_{i=1}^N \epsilon_i \ A_i,
\end{equation} 
where $\epsilon_i$ are ellipticities of the individual iso-intensity 
contours measured as stated earlier and $M$ is the multiplicity at that 
level. The symbols $A_i$ and $A_{max}$ have similar meanings as before. 

To construct ${\epsilon}_{agg}$, we take the union of all components 
present at a given level and form a collective region. The integrated 
region can be expressed as 
\begin{equation} 
R = R_1 \cup R_2 \cup \cdot \cdot \cdot \cup R_N, 
\end{equation} 
where $R_i$ is the region enclosed by each component. Subsequently we 
find the components of the area tensor and the ``auxiliary ellipse'' for 
the region $R$.

It is worth mentioning that the behavior of ${\epsilon}_{agg}$ is 
similar to conventional ellipticity measure based on inertia tensor 
(Carter \& Metcalf 1980). But the construction procedure of these two 
measures are different. The conventional method finds the eigenvalue 
of the inertia tensor for an annular region enclosing mass density or 
surface brightness, on the other hand, the method based on MFs finds 
the eigenvalues of regions enclosed by the contour(s) is assumed to be 
homogeneous.   

We have expressed ellipticities after averaging the estimates at all 
density/brightness levels. Our final result is, therefore, expressed by 
$\bar{\epsilon}_{eff}$ and $\bar{\epsilon}_{agg}$ rather than 
$\epsilon_{eff}$ and $\epsilon_{agg}$.

\subsection{Toy Models}
\begin{figure}
\epsscale{1.1}
\plotone{\figdir/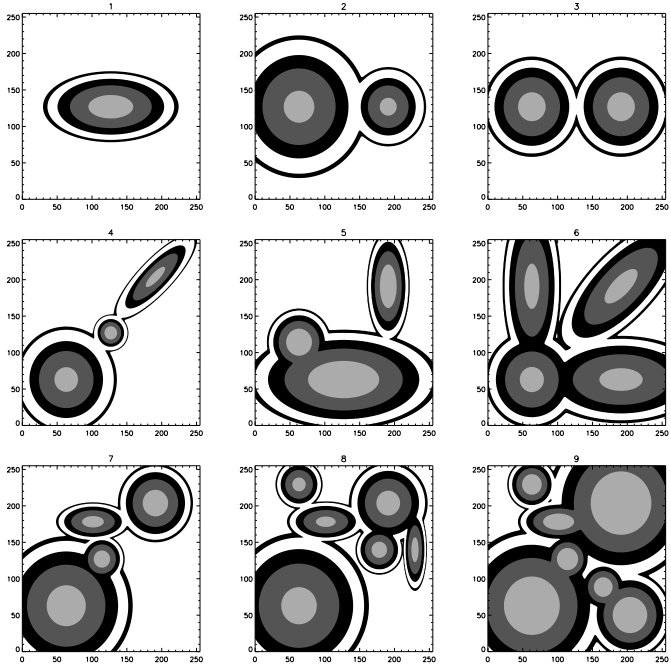}
\caption{Contour plots of toy clusters at different brightness levels. 
The choice of levels is arbitrary. The multi-modal clusters have clumps 
with different peak brightness. For all clusters the outer line represents 
the percolation level at which substructures merge and form 
a single, large system.\label{toy_img}}
\end{figure}

To get a better feeling of the parameters mentioned above we provide 
an illustrative example with toy models. One can think of these toy 
models as snap shots of different X-ray clusters in projection taken at 
one particular time. We include clusters with different types of internal 
structures in Fig. \ref{toy_img}: unimodal elliptic structure (panel 1), 
asymmetric and symmetric bimodal clusters (panels 2 and 3, respectively), 
cluster with filamentary structure (panel 4) etc. The multi-modal clusters 
have clumps with different peak brightness. We show contour plots of toy 
models at different brightness levels where the choice of levels is 
arbitrary. For all clusters the outer line represents the percolation 
level at which substructures merge with one another forming a single, 
large system. 

\begin{figure}
\epsscale{1.12}
\plotone{\figdir/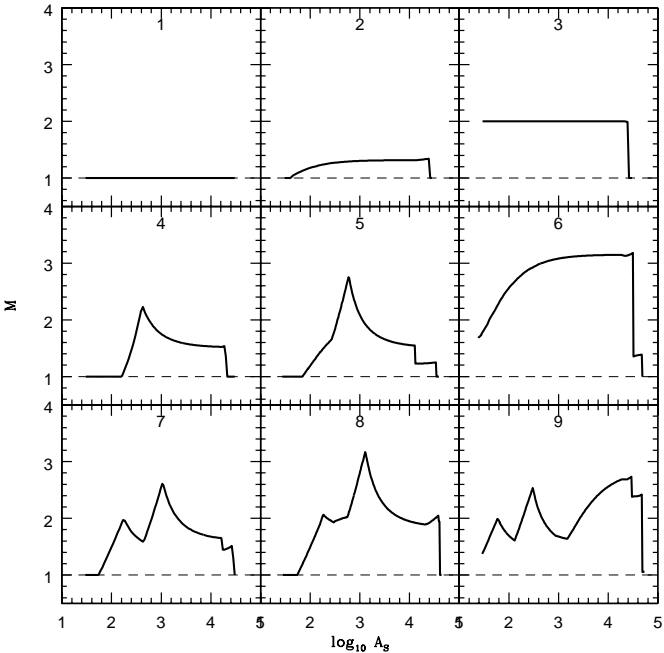}
\caption{Multiplicity as a function of area for toy models. The number 
at  each panel corresponds to the cluster number shown in Fig. \ref{toy_img}. 
For details see text. \label{toy_mul}}
\end{figure}

Multiplicity as a function of area (in grid unit) is shown in  Fig. 
\ref{toy_mul}. As mentioned earlier $M$ is sensitive to the sizes of 
the substructures. The simplest case to check this is to take a bimodal 
cluster. 
For a bimodal structure with unequal sized sub-clumps (panel 2), the 
fractional value of multiplicity ($1 < M < 2$) tells us that the 
components of the system have different sizes. The isolated components 
eventually percolate giving $M = 1$ at low brightness level, i. e, at 
larger area. On the other hand, for a cluster with equal 
components $M = 2$ until percolation occurs (panel 3). For clusters 
with three components (panels 4 and 5), we see that for a short range 
of brightness levels the components are well separated where two of 
these are bigger then the third one ($2 < M < 3$). Afterwards two of 
the three clumps merge giving $1 < M < 2$. These two remaining 
components eventually percolate to become a single system. The clumps 
in panel 6 are distributed around the center. For this cluster, we see 
two unequal size but well separated clumps ($1 < M < 2$) with same peak 
brightness. The behavior of clusters in panels 7 and 8 is similar except 
that they have different number of substructures. The cluster in panel 
9 has the largest number of components (a total of 7). Two of its clumps 
are so large compared to the other ones that they dominant mostly. The 
multiplicity is always in the range $1 < M < 3$ reflecting the merging 
of clumps at different levels. 

\begin{figure}
\epsscale{1.12}
\plotone{\figdir/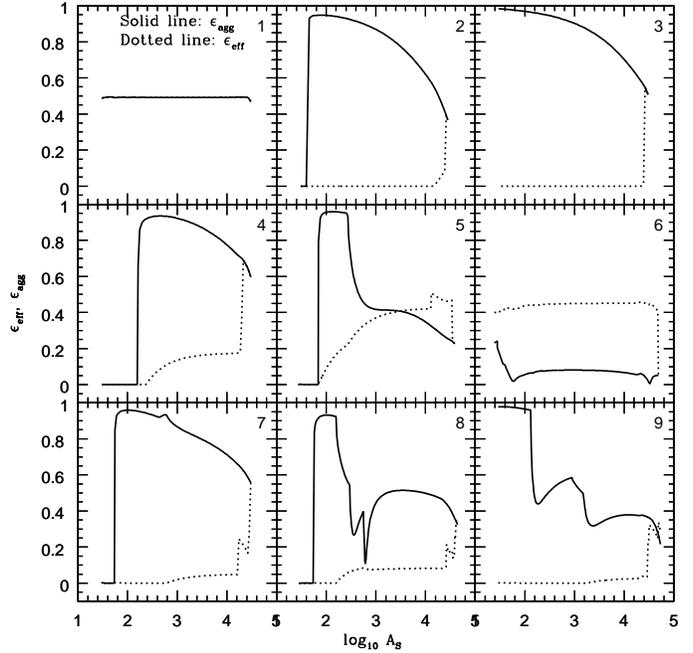}
\caption{Ellipticity as a function of area for toy models. The number 
at  each panel corresponds to the cluster number shown in Fig. \ref{toy_img}. 
The solid and dotted line represent, respectively, $\epsilon_{agg}$ 
and $\epsilon_{eff}$. For details see text. \label{toy_ell}}
\end{figure}

Ellipticity as a function of area is shown in  Fig. \ref{toy_ell}. 
In this figure the solid and dotted line represent, respectively, 
$\epsilon_{agg}$ and $\epsilon_{eff}$. 
For the unimodal cluster in panel 1, $\epsilon_{eff} = \epsilon_{agg}$.
For the bimodal cluster in panel 2, the estimate of $\epsilon_{eff}$ 
gives an ellipticity weighted more by the larger component. In this case 
it is zero. 
However, for a bimodal system with equal sized sub-clumps but different 
elongation, $\epsilon_{eff}$ will give an average elongation of the 
two. The estimate of ${\epsilon}_{agg}$, on the other hand, tells 
us about the overall shape of that system irrespective of the sizes 
and elongations of its sub-clumps. Due to the presence of two isolated 
components, the system itself appears more elongated than the shape 
of its sub-clumps. 

The important point to note that the estimate provided by ${\epsilon}_{agg}$ 
depends not only on the relative sizes of the components but also on their 
relative separation. This is reflected in all panels containing multi-clump 
clusters. With the decrease of brightness, as the clumps get bigger and 
appear close to one another, ${\epsilon}_{agg}$ gets smaller. 

For a multi-component system with filamentary structure, 
$\epsilon_{eff} < \epsilon_{agg}$ (panel 4). If components are 
distributed around the cluster center, $\epsilon_{eff} > \epsilon_{agg}$ 
(panel 6). The cluster in panel 5 has the unique property that is 
shown separately by clusters in panel 4 and 6. In transition from 
peak brightness to lower level, the cluster changes its filamentary 
shape to the one where the components are distributed over a region 
around the center. The $\epsilon_{agg}$ profile in panel 8 shows that 
in the range, $2.2 < \log_{10} A_S < 2.8$, the cluster develops two, 
almost equal size clumps that are very close to each other. Cluster 
in panel 7 follows the behavior of a bimodal cluster except that there 
is jump in between $2.6 < \log_{10} A_S < 2.8$ where the cluster changes 
its structure having two unequal size clumps to two equal size clumps.
The shape of the cluster in panel 9 changes consistently following the 
merging of its clumps at different brightness.

\subsection{Example of Simulated Clusters} 
\begin{figure}
\epsscale{1.15}
\plotone{\figdir/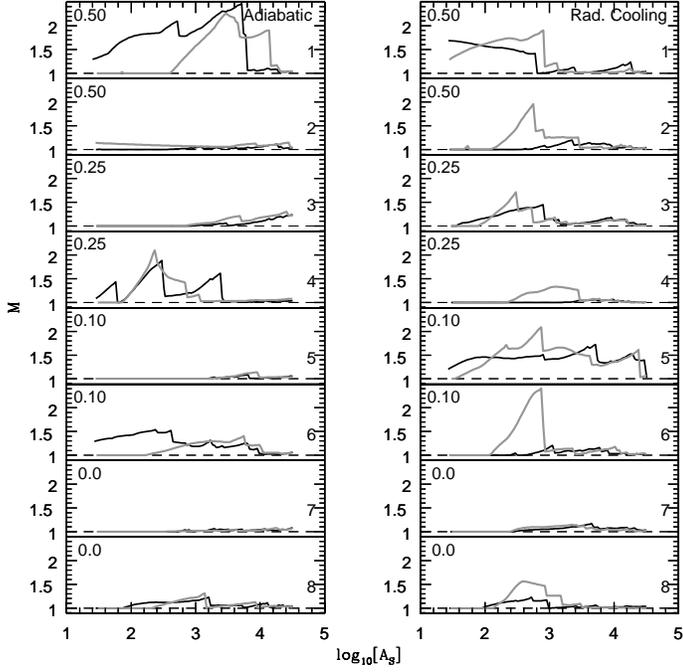}
\caption{The multiplicity (${M}$) as a function of area ($A_S$) for a 
selection of clusters at $z = 0.50, 0.25,$ $0.10,$ and $0.0$. Two 
clusters from each redshift are shown. The dark and gray solid lines 
represent the matter and X-ray clusters, respectively. For a uni-modal 
cluster, $M = 1$, on the other hand for a multi-modal cluster, $M > 1$. 
Figure shows that multiplicity is, in general, greater than 1 in the 
entire redshift range for clusters simulated with radiative cooling 
(right panels) indicating a slower evolution than in the adiabatic 
sample (left panels). \label{multip_area}}
\end{figure}

\begin{figure}
\epsscale{1.15}
\plotone{\figdir/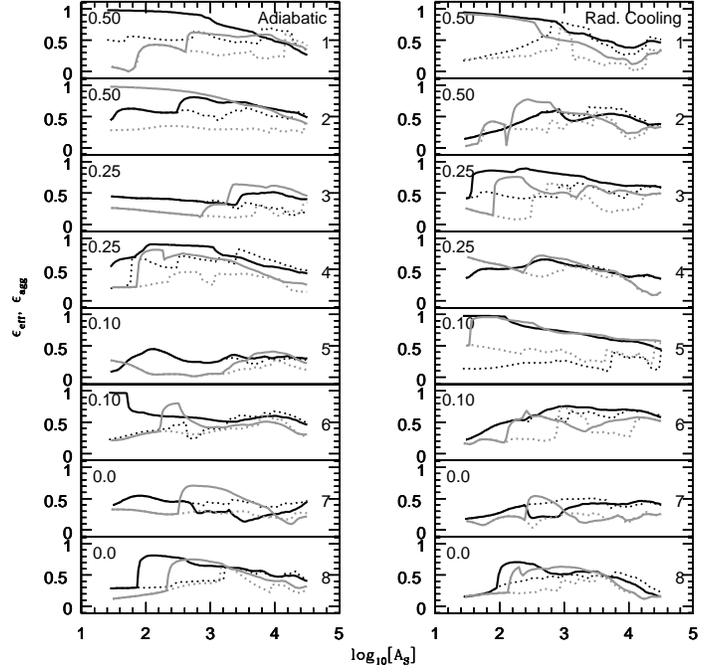}
\caption{The effective (${\epsilon}_{eff}$) and aggregate 
(${\epsilon}_{agg}$) ellipticity as a function of area ($A_S$) for 
the same clusters as in Fig. \ref{multip_area}. Dotted and solid lines 
are used for ${\epsilon}_{eff}$ and ${\epsilon}_{agg}$, respectively. 
The matter and X-ray clusters are shown, respectively, by the dark 
and gray lines. In most cases the central part of clusters consists 
of single peak (i. e., ${\epsilon}_{eff} = {\epsilon}_{agg}$). 
The central region of these clusters do not appear spherical. Rather 
the region appears to have some degree of flattening. Multi-peak 
systems, mostly bimodal clusters with un-equal size sub-clumps 
(${\epsilon}_{eff} < {\epsilon}_{agg}$), however, are not uncommon. 
At low brightness levels, i. e., in the outer regions of cluster, 
the sub-clumps appear to be homogeneously distributed 
(${\epsilon}_{eff} > {\epsilon}_{agg}$). \label{ellip_area}}
\end{figure}

We demonstrate the behaviors of $M$ and the variants of $\epsilon$ as a 
function of area for several simulated clusters (Figs. \ref{multip_area} 
and \ref{ellip_area}). For each sample we choose two clusters at each 
redshift for illustration purpose. We use dark and gray lines to 
represent matter and X-ray clusters, respectively. Fig. \ref{multip_area} 
shows that both matter and X-ray clusters with cooling have higher 
number of sub-clumps than those without cooling. 
Fig. \ref{ellip_area} shows that in most cases the central part of 
clusters consists of single peak (${\epsilon}_{eff} = {\epsilon}_{agg}$). 
The central region of these clusters do not appear spherical. Rather 
the region appears to have some degree of flattening. We see that 
multi-peak systems, mostly bimodal clusters with un-equal size 
sub-clumps (${\epsilon}_{eff} < {\epsilon}_{agg}$), are not uncommon 
for these clusters. In low brightness levels, i. e., in the outer 
regions of cluster, the sub-clumps appear to be homogeneously 
distributed (${\epsilon}_{eff} > {\epsilon}_{agg}$). 

\section{Results}
The objective of this paper is to study morphological evolution in 
simulated cluster using $M$ and $\epsilon$ as quantitative measures. 
The parameters represent the shape 
characteristics of a set of iso-intensity contours corresponding to a 
set of density/brightness levels. The levels represent equal interval 
in area (i.e. size) in log space. 
We select the range of area in between $30 \sim 35000$ (in grid unit) 
with 100 intervals. The lower limit is to avoid the discreteness of the 
grid whereas the upper limit covers the significant part of the entire 
image of each cluster analyzed. In fact with this range we can 
analyze clusters from $\sim 80$ h$^{-1}$ kpc up to a distance $\sim 2$ 
h$^{-1}$ Mpc. We do not smooth either the cluster image or its 
iso-intensity contours.

\subsection{Comparison between Cluster Samples}
\begin{figure}
\epsscale{1.15}
\plotone{\figdir/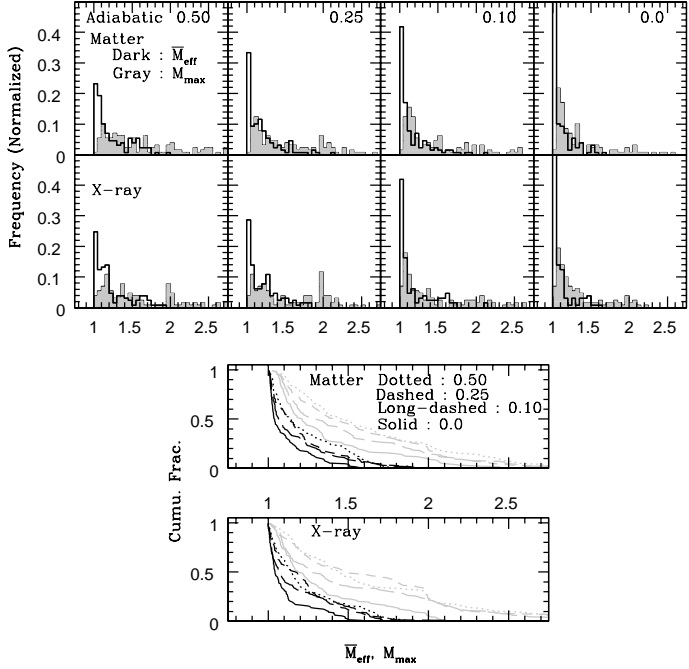}
\caption{Multiplicity in adiabatic sample. The figure shows the 
normalized frequency (probability distribution function; top two 
panels) and the cumulative fraction (cumulative distribution function; 
bottom two panels) of $\bar{M}_{eff}$ and ${M}_{max}$ at different 
redshifts for matter and X-ray clusters. The total number of cluster 
in this sample is, $N_{ad} =  129$. The numbers at the top-right in 
the uppermost panels show the redshifts. \label{adiabatic_m}}
\end{figure}

\begin{figure}
\epsscale{1.15}
\plotone{\figdir/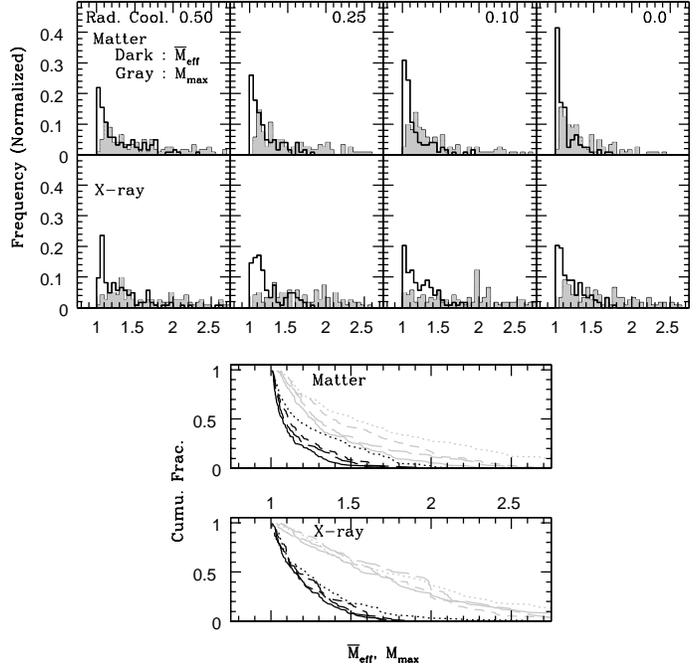}
\caption{Multiplicity in radiative cooling sample. The normalized 
frequency (top two panels) and the cumulative fraction (bottom two 
panels) of $\bar{M}_{eff}$ and ${M}_{max}$ at different redshifts 
for matter and X-ray clusters. The total number of cluster in this 
sample is, $N_{rc} =  123$. \label{radiative_m}}
\end{figure}

\begin{figure}
\epsscale{1.15}
\plotone{\figdir/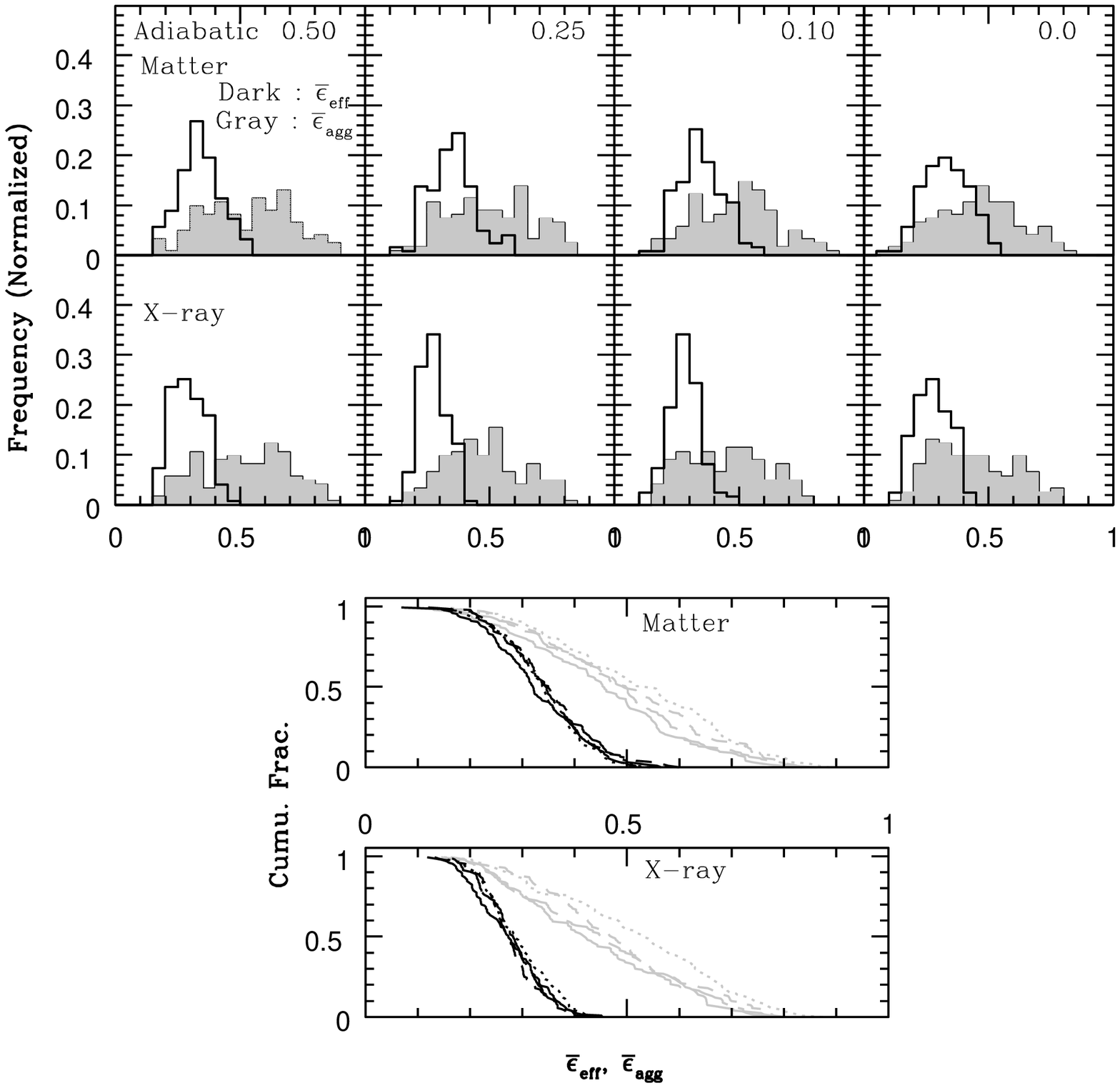}
\caption{Ellipticity in adiabatic sample. The normalized frequency 
(top two panels) and the cumulative fraction (bottom two panels) 
of $\bar{\epsilon}_{eff}$ and $\bar{\epsilon}_{agg}$ at different 
redshifts for matter and X-ray clusters. \label{adiabatic_e}}
\end{figure}

\begin{figure}
\epsscale{1.15}
\plotone{\figdir/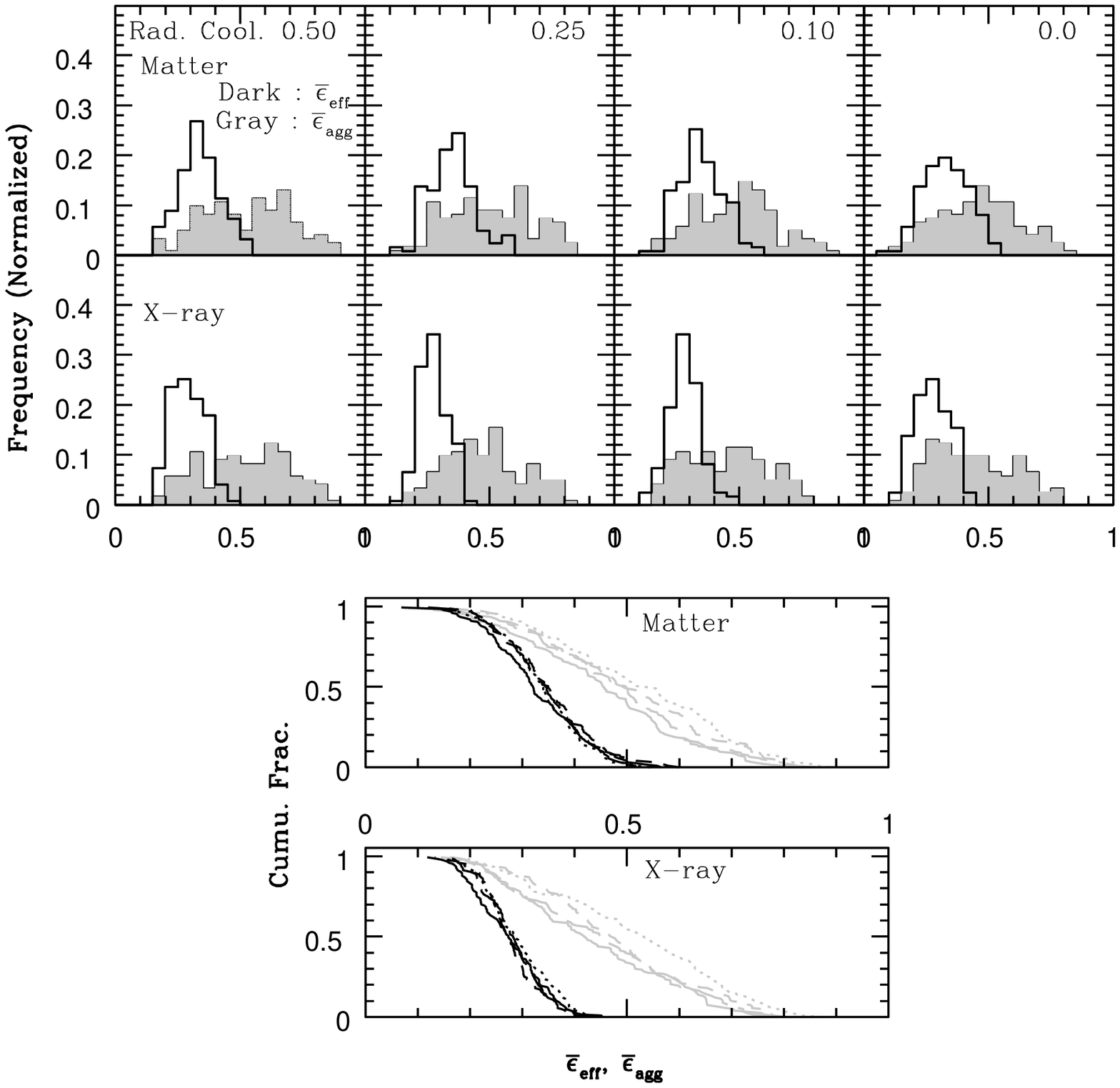}
\caption{Ellipticity in radiative cooling sample. The normalized 
frequency (top two panels) and the cumulative fraction (bottom two 
panels) of $\bar{\epsilon}_{eff}$ and $\bar{\epsilon}_{agg}$ at 
different redshifts for matter and X-ray clusters. \label{radiative_e}}
\end{figure}

The probability and cumulative distribution functions of these parameters 
determined from the cluster samples are shown in Figs. \ref{adiabatic_m} 
to \ref{radiative_e}. In each of these figures the probability function 
(i. e., frequency normalized by the number of elements within the sample) 
are shown at the top two panels at 4 different redshifts. The cumulative 
functions are shown at the bottom two panels where dotted, dashed, long 
dashed, and solid lines are used for the cumulative functions derived at 
$z = 0.0, 0.10, 0.25,$ and $0.50$, respectively. In these figures dark 
and gray lines represent, respectively, $\bar{M}_{eff}$ and $M_{max}$; 
similarly dark and gray lines represent, respectively, 
$\bar{\epsilon}_{eff}$ and $\bar{\epsilon}_{agg}$. As mentioned earlier 
these parameters provide estimate averaged over 100 brightness levels. 

From a visual examination of the distribution functions in Figs. 
\ref{adiabatic_m} to \ref{radiative_e} one can notice an evolutionary 
trend in cluster morphology since the gross properties of clusters 
indeed change with redshifts. Our goal is to determine the statistical 
significance of this trend and than compare it with the observation. 
We should mention that multiplicity as defined here had not been used 
before. It is, therefore, not possible to make a direct comparison with 
other studies or observations. As a result we will use only ellipticity 
for comparison with observations. We will explore the robustness and 
sensitivity of $M$ in our future work on Sloan Digital Sky Survey (SDSS) 
clusters.

To check whether or not the trend is significant, we have performed K-S 
test between cumulative functions at two different $z$ for each parameter. 
The test is repeated for all combinations of redshifts and the results are 
presented in the tables 1 to 4 where d and p represent the K-S statistic 
and significance level probability. The tables show test results only for 
those redshift combinations that may indicate a significant evolution, 
i. e., only those results for which p $< 0.1$. Recall that a small value 
of p corresponds to a significant difference between two distribution 
functions.

In terms of $M$, the test results give a strong indication of evolution 
when cumulative functions at different $z$ are compared 
(Figs. \ref{adiabatic_m} and \ref{radiative_m}; tables 1 and 2).  
The distribution functions of $\epsilon$, on the other hand, show slow 
or very weak evolution. (Figs. \ref{adiabatic_e} and \ref{radiative_e}; 
tables 3 and 4). This is a general outcome for both samples of clusters.

Figs. \ref{adiabatic_m} and \ref{radiative_m} show that clusters in the 
cooling sample have a higher value of multiplicity at all redshifts 
compared to those in the adiabatic sample. The low abundance of single 
component systems with radiative cooling indicates that the dense, cool 
core substructures are long lived features (Motl et al. 2003).

Table 1 shows the K-S tests of the distributions of $M$ and $M_{max}$ for 
adiabatic clusters. From the table we find that these clusters, in general, 
show marked decline in sub-clumps from $z = 0.50, 0.25$ to $z = 0.0$ 
(p $\simeq 10^{-4}$). In between $z = 0.10$ and $z = 0.0$, the  evolution 
is less significant (p $\simeq 10^{-2}$).

Table 2 shows that the evolution of dark matter clusters in the cooling 
sample is stronger compared to the X-rays.
The low significance is a reflection of less efficient merging in the 
X-ray gas. In fact in terms of $M_{max}$, the X-ray clusters do not 
show any significant change in the distribution functions over redshift. 
For dark matter, the distribution functions of this parameter do 
evolve from $z = 0.50, 0.25$ to $z = 0.0$.

The distribution functions of ellipticities for both samples are shown 
in Figs. \ref{adiabatic_e} and \ref{radiative_e}. We see that the X-ray 
clusters, in general, are more regular than those of dark matter. The 
X-ray gas evolves in the dark matter potential and due to the radiation 
pressure it is distributed homogeneously in the background. As a result 
the morphological characteristics of systems containing baryonic fluids 
will be more regular. 
For these clusters the irregular components are distributed over a region 
instead of concentrating along one direction making them appear regular. 
A comparison of $\bar{\epsilon}_{eff}$ with $\bar{\epsilon}_{agg}$ for 
dark matter clusters shows that the sub-clumps are not distributed 
homogeneously around the central region. Rather these clumps are spread 
out mostly in one direction forming filamentary structure, as indicated by 
the larger value of $\bar{\epsilon}_{agg}$.

No significant evolution is signaled by $\bar{\epsilon}_{eff}$ 
either for matter or for X-ray clusters in the adiabatic sample (table 3). 
It indicates that the shape of individual components of a cluster at one 
redshift may appear similar at other redshifts. 
Since $\bar{\epsilon}_{eff}$ puts emphasize on individual components, 
therefore, it is not unusual to see no evolution quantified by this 
parameter. In terms of $\bar{\epsilon}_{agg}$, we see that the overall 
shapes of X-ray clusters do evolve from $z = 0.50$, $0.25$ towards 
$z = 0.0$ (p $\simeq 10^{-5}$). The matter clusters, however, show less 
significant evolution (p $\simeq 10^{-2}$) from $z = 0.50$ to $z = 0.0$. 

Once again no evolution is signaled by $\bar{\epsilon}_{eff}$ for 
clusters with radiative cooling (table 4). We see a moderate 
evolutionary trend is given by $\bar{\epsilon}_{agg}$ from $z = 0.50$ 
to $z=0.10$ and $0.0$ with p $\simeq 10^{-3}$. 

As a final check with the distribution functions, we take projection 
along each axis at a time and repeat the K-S tests. Recall that in this 
case sub-samples (along each axis) have 41 clusters in the adiabatic limit 
and 43 clusters with radiative cooling. The results of these sub-samples 
show insignificant variation from the bigger primary sample. Therefore, it 
is less likely that the overall result is affected by projection effect.

\begin{figure}
\epsscale{1.15}
\plotone{\figdir/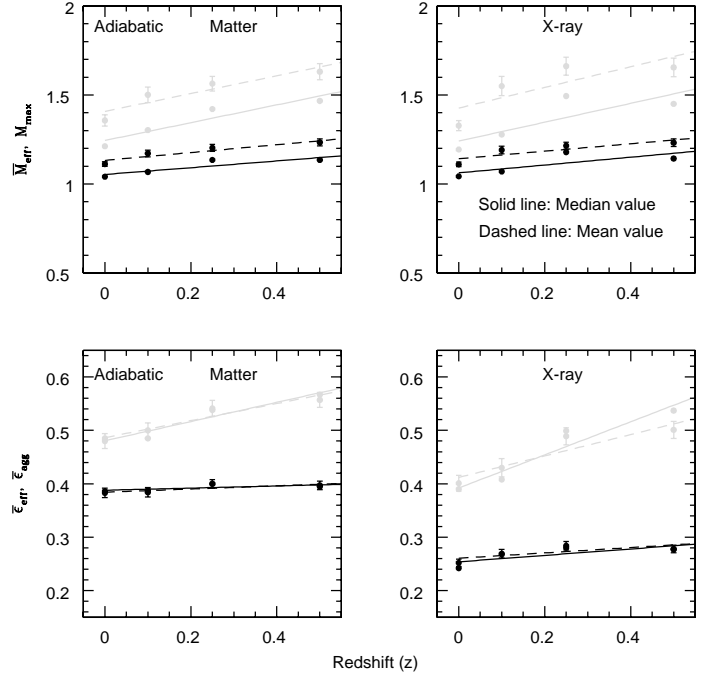}
\caption{Adiabatic sample: Redshift ($z$) evolution of $M$ and $\epsilon$ 
for matter and X-ray clusters. At each redshift, the figure shows median, 
and mean with 1$\sigma$ error bar derived from the probability distribution 
functions. The solid and dashed lines are the best fit lines, respectively, 
for median and mean (see table 5). The top panels show multiplicity where 
dark and gray lines are used for $\bar{M}_{eff}$ and $M_{max}$, respectively. 
The bottom panels show ellipticity where dark and gray lines are used for 
$\epsilon_{eff}$ and $\epsilon_{agg}$, respectively. \label{adiabatic_evol}}
\end{figure}

\begin{figure}
\epsscale{1.15}
\plotone{\figdir/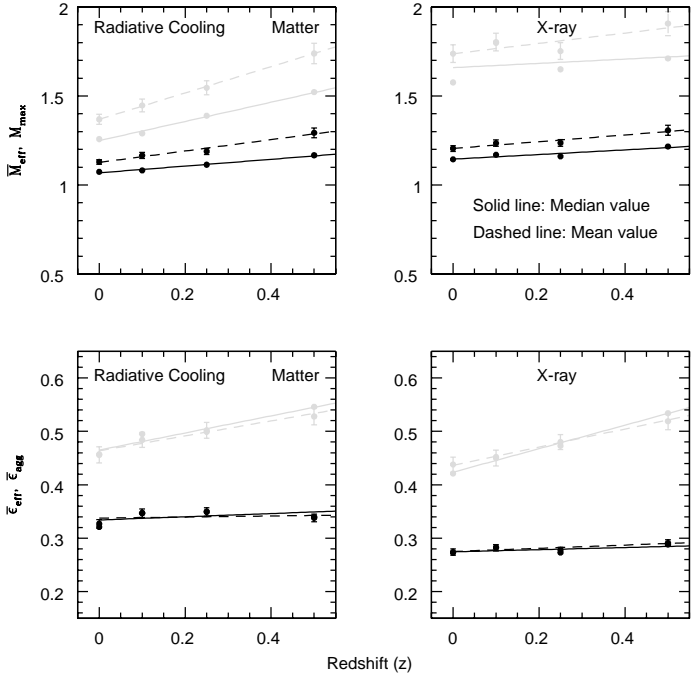}
\caption{Radiative cooling sample: Redshift ($z$) evolution of $M$ and 
${\epsilon}$ for matter and X-ray clusters. Presentation style is similar 
to Fig. \ref{adiabatic_evol} except that this time table 6 is used to draw 
the best fit lines. \label{radiative_evol}}
\end{figure}

We present redshift evolution of the measures for adiabatic and cooling 
sample in Figs. \ref{adiabatic_evol} and \ref{radiative_evol}, 
respectively. At each redshift we show the median and the mean with 
1$\sigma$ error bar derived from the distribution functions. 
We quantify the rate of evolution by the slope of the line obtained from 
the best least-squares fit to the $M$ vs. $z$ and $\epsilon$ vs. $z$ 
relationship shown in these figures where rate means either $d M/d z$ or 
$d \epsilon/d z$. The solid and dashed lines represent the best fit to the 
data given, respectively, by the median and the mean. We summarize our 
measurements in tables 5 and 6. 

Several conclusions can be drawn from these tables. 
First, the dark matter shows very similar evolution in both samples 
of clusters. Second, the X-ray clusters in adiabatic simulation evolve 
faster than those with radiative cooling. Third, the rate of ellipticity 
evolution is higher for X-ray clusters irrespective of cooling history. 
Fourth, as far as the strength of evolution is concerned, multiplicity, 
especially $M_{max}$, appears to be more sensitive indicator than 
ellipticity. 

We note that the measured quantities for the dark matter in the adiabatic 
and cooling cluster samples are very similar. This is a check on the 
consistency of the simulations and analysis. The result is expected as 
the N-body segment of the simulations are identical in the two cluster 
samples with the exception of the gas that has relatively minor 
contribution to the total gravitational potential.
The LSS of adiabatic and cooling clusters are generally similar but their 
small scale structures are dictated by the overall cluster properties 
rather than perturbative interactions (Motl et al. 2003). In adiabatic 
clusters infalling sub-clumps mix into the main cluster medium in a faster 
manner relative to the radiative cooling where sub-clusters can be long 
lived. This is the reason behind fast evolution of adiabatic X-ray clusters. 
The relaxation times scale for collisionless particles is much longer 
than that of the collisional gas particles (Frenk et al. 1999; Valdarnini, 
Ghizzardi \& Bomometto 1999). Therefore matter clusters will appear not 
only more elongated than the X-ray gas distribution but the redshift 
evolution of their shapes will also be slow. More spherical configuration 
for X-ray clusters is also expected from the point of view that intracluster 
gas is in hydrostatic equilibrium whereas the dark matter is distributed 
like galaxies (Kolokotronis et al. 2001).

FMM have analyzed the same sets of simulated clusters in the redshift 
range $0.0 - 0.25$ using eccentricity ($e$) as the quantitative measure. 
The eccentricity is defined as, $e = \sqrt{1 - \lambda_2/\lambda_1}$ 
where $\lambda_1$ and $\lambda_2$ are eigenvalues of the inertia tensor 
($\lambda_1 > \lambda_2$). Note that the eigenvalues are proportional to 
the square of the semi-axes of an ellipse. FMM used moment of inertia 
technique for an annular region to calculate $e$. The size of the annulus 
is fixed either by r$_{outer}$ or r$_{200}$ where r$_{outer}$ is the 1.5 
$h^{-1}$ Mpc aperture and the r$_{200}$ is the radius from cluster center 
up to a distance where the density is $\sim$ 200 times the background. 

To make a comparison with FMM, we measure eccentricity for our data sets. 
Using both effective ($\bar{e}_{eff}$) and aggregate ($\bar{e}_{agg}$) 
eccentricities, we find that 1) the results agree, at least, qualitatively 
in the slow evolution of simulated clusters and 2) both studies show that 
clusters without cooling evolve faster compared to those having radiative 
cooling in a $\Lambda$ dominated universe. 

However, we note two systematic differences between the results. First, 
our results show that matter distributions generally are more elongated 
than the X-ray gas whereas in FMM both of these systems have comparable 
eccentricities. As mentioned earlier the collisionless dark matter 
distribution relaxed in longer time than the hot gas, it seems unlikely 
that both systems will have comparable flattening within the same 
cosmological time slot (i. e., $0 < z < 0.5$). 
Second, at each redshift our estimates are higher compared to those of 
FMM. FMM focuses mainly to determine the shape in the outer region of 
galaxy-clusters. 
Our method, on the other hand, measures shapes from the central regions 
towards the edge of clusters (up to $\sim 2$ h$^{-1}$ Mpc). 
In cluster morphology it is important to know which part of a cluster 
needs to be analyzed that will give a realistic estimate of its shape. 
For example, to study clusters with cool core one should include the 
central region in the analysis to see to what extent it influences the 
overall shape since cooling is is a small scale phenomenon that occurs 
within 100 to 200 h$^{-1}$ kpc of cluster center.
Emphasizing only on the outer part for cool core clusters, therefore, 
may not reflect the estimate of actual shape. 
Clusters with or without cooling appear to be more elongated around the 
central region (see Fig. \ref{ellip_area}; note that a similar trend in 
simulated cluster without cooling has also been reported by Jing \& Suto 
2002). Since we estimate an average eccentricity, our result is weighted 
more by the large flattening around the central part. The difference in 
results is, therefore, a reflection of different methodology.

\subsection{Comparison with Observed Clusters}
The optical sample of MCM has 138 ACO clusters with $z < 0.1$. It has 
been compiled from West \& Bothun (1990); Rhee, van Haarlem \& Katgert 
(1991) and Kolokotronis et al. (2001). We do not see any evolution for 
this sample ($d \epsilon/d z \sim 0.03$). Plionis (2002) has the largest 
sample of optical clusters with measured ellipticity. It has 407 APM 
clusters with $z < 0.18$. For our purpose this sample is slightly 
better than the MCM since it has clusters with higher redshift. Besides, 
$\sim 30$ clusters of this sample are also present in the MCM. The rate 
of evolution for this sample is $d \epsilon/d z \sim 0.7$. 
However, if both are combined, replacing the common ones by the 
APM clusters, the rate increases. The combined sample of $\sim 500$ 
optical clusters with $z < 0.18$ shows $d \epsilon/d z \sim 1.06$. 

It is rare to find a large sample of X-ray clusters with up-to-date 
ellipticity measure. We use the sample of MCM that is compiled from 
Mcmillan, Kowalski \& Ulmer (1989) and Kolokotronis et al. (2001). It 
only has 48 clusters with $z < 0.1$ which is three times smaller than 
the MCM optical sample and an order of magnitude smaller than the APM. 
It also has low redshift limit than the APM. The rate of 
evolution for this sample is $d \epsilon/d z \sim 1.7$. The result 
suggests faster evolution for X-ray clusters than the opticals.
Although the result is in accordance with our expectation, we should 
not make any definite conclusion because of the size of the sample. A 
large sample of X-ray clusters with better selection criteria and 
extended over wide redshift is needed to be conclusive.
 
We reanalyze the APM clusters and the combined sample imposing redshift 
cutoff $z < 0.1$ in order to have redshift range consistent with the MCM 
X-ray sample. These samples show, respectively, $d \epsilon / d z \sim 1.02$ 
and $\sim 1.0$. Even though evolution of optical clusters gets faster in 
this redshift range, it is still slower than that of the X-rays.
 
When we compare the rate of evolution between simulated and observed 
cluster ellipticity, we find slower evolution in simulations. In spite 
of this, it is interesting that the simulation is able to capture the 
essence of reality: faster evolution of gas than the dark matter 
distribution.
\section{Conclusions}
Numerical simulations provide an unique opportunity to understand 
the underlying physics of structure formation. In order to be 
representative of the real world, the results from simulations should 
agree with observations. Observations provide evidence of 
morphological evolution in galaxy-clusters (Melott, Chambers \& Miller 
2001; Plionis 2002; Jeltema et al. 2003), simulations should show 
similar evolution. 
With this in mind, we have studied redshift evolution of cluster 
morphology simulated, respectively, in the adiabatic limit and with 
radiative cooling. 

Since observed clusters are projected along the line of sight and lack 
of full 3 dimensional information we, therefore, use projected simulated 
clusters. Each cluster image is within a 8 h$^{-1}$ Mpc frame containing 
360 $\times$ 360 pixels. The clusters are analyzed at 100 different 
density/brightness levels using multiplicity and ellipticity as two 
different probes to quantify morphological evolution. 

Our results show that if cluster shapes are quantified by ellipticity, 
than simulations do not model the observed local universe well enough. 
The results indicate that the simulated clusters do evolve with redshift 
but the evolution is slower than the observed one. The outcome of our 
analyses, however, should be taken with a caution since the measurements 
from different samples do not agree to be conclusive on the slower 
evolution. 
Take, for example, the optical clusters for $z < 0.1$: the APM sample 
shows $d \epsilon / d z \sim 1.02$, whereas the MCM sample shows 
$d \epsilon / d z \sim 0.03$. 
A recent study by Flin, Krywult \& Biernacka (2004; hereafter FKB) 
reports very weak evolution for a sample of 246 ACO clusters for
redshift range $0 < z < 0.3$. They analyze clusters at five different 
annular radii and the mean of these estimates shows 
$d \epsilon / d z \sim 0.013$. For $z < 0.1$, their result also indicates 
weak evolution (Flin, 2004).
In terms of ellipticity, therefore, we can see clearly from table 5 and 
6 that the evolution is slower for both sets of simulations with respect 
to the APM sample but it is stronger compared to both MCM and FKB. 
   
A preliminary analysis of a sample of 800 clusters constructed from the 
SDSS shows that ellipticity evolution of optical clusters within 
$z < 0.1$ is weaker than that of the APM clusters. The result also shows 
that clusters with different mass limits evolves differently. Large, 
massive clusters ($M \sim 10 ^{15} M_{\odot}$) have stronger evolution 
compared to the less massive clusters 
($M \sim 10 ^{13} - 10 ^{14} M_{\odot}$) (Miller 2004). 
The sample is uniform with a well documented selection function and high 
degree of completeness. If this is the case then we can infer that the 
cluster samples mentioned above have less uniformity in mass range: the 
APM catalogue is biased towards the massive cluster and both MCM and FKB 
samples contain less massive clusters. Note that we have checked the 
evolution in simulated clusters with the mass limits as mentioned above 
but have not found any difference.    

The discrepancy in the optical samples is an indication of different 
selection criteria used to construct the catalogues. Larger and more 
complete catalogues obtained from Sloan Digital Sky Survey and XMM-Newton 
survey may be able to shed more light into this issue.
It is also likely that numerical simulations may lack crucial physics that 
needs to be included (see FMM for discussion). In the forthcoming paper in 
this series we will analyze cluster samples simulated with various gas 
physics, for example, star formation and feedback from supernovae explosion. 
The result of that study may give some clue to gain better insight of the 
discrepancy.

Finally, our result indicates that multiplicity ($M_{max}$) indicates 
relatively stronger evolution in cluster morphology compared to ellipticity 
This particular result suggests that it is a more sensitive parameter 
than ellipticity (see also Thomas et al. 1998; Suwa et al. 2003). We will 
further explore the sensitivity of this measure in our future work on 
simulation and observed clusters.  
\\
\\
\noindent{\bf Acknowledgments}
We thank M. Plionis for providing the APM cluster ellipticity data. 
We also thank Scott W. Chambers for the MCM data sets. NR thanks Hume 
Feldman, Bruce Twarog, Brian Thomas, and Stephen Floor for many useful 
discussions. NR acknowledges the GRF support from the University of Kansas 
in 2003-2004. ALM acknowledges the financial support of the US National 
Science Foundation under grant number AST-0070702.

\begin{table}
\caption{Multiplicity in the adiabatic sample. 
The K-S test result between cumulative distribution functions at 
different redshifts for $\bar{M}_{eff}$ (top) and $M_{max}$ (bottom). 
In the tables $z$, d, and p stand 
for redshift, K-S statistics, and the significance level probability, 
respectively. In K-S test, a small value of the probability suggests 
that two distributions are different from each other.}
\begin{tabular}{|c|c|c|c|c|c|}   \hline
{$z$}  &\multicolumn{2}{c|} {Matter} 

       &\multicolumn{2}{c|}{X-ray}   \\ \cline{2-5}

                   	   &{d}
           		   &{p}     
 			   
                   	   &{d}
           		   &{p}	     \\ \cline{1-5} 		
		
0.00 - 0.10   &0.17  &4.1 $\times 10^{-2}$  &0.18  &2.9 $\times 10^{-2}$ 
\\ \cline{1-5}
0.00 - 0.25   &0.26  &3.3 $\times 10^{-4}$  &0.33  &7.3 $\times 10^{-7}$ 
\\ \cline{1-5}
0.00 - 0.50   &0.33  &1.5 $\times 10^{-6}$  &0.34  &3.7 $\times 10^{-7}$ 
\\ \cline{1-5}
0.10 - 0.25   &0.18  &2.9 $\times 10^{-2}$  &0.18  &2.9 $\times 10^{-2}$ 
\\ \cline{1-5}
0.10 - 0.50   &0.22  &3.7 $\times 10^{-3}$  &0.22  &3.7 $\times 10^{-3}$ 
\\ \cline{1-5}	
0.25 - 0.50   &-     &-                     &-     &-      \\ \hline \hline

{$z$}  &\multicolumn{2}{c|} {Matter} 

       &\multicolumn{2}{c|}{X-ray}   \\ \cline{2-5}

                   	   &{d}
           		   &{p}     
 			   
                   	   &{d}
           		   &{p}	    \\ \cline{1-5} 		
0.00 - 0.10   &0.17  &4.1 $\times 10^{-2}$  &0.17  &4.1 $\times 10^{-2}$ 
\\ \cline{1-5}
0.00 - 0.25   &0.26  &3.3 $\times 10^{-4}$  &0.30  &1.9 $\times 10^{-5}$ 
\\ \cline{1-5}		
0.00 - 0.50   &0.30  &1.0 $\times 10^{-5}$  &0.30  &1.0 $\times 10^{-5}$ 
\\ \cline{1-5}		
0.10 - 0.25   &-     &-                     &0.19  &2.0 $\times 10^{-2}$ 
\\ \cline{1-5}
0.10 - 0.50   &0.22  &2.4 $\times 10^{-3}$  &0.20  &8.9 $\times 10^{-3}$ 
\\ \cline{1-5}		
0.25 - 0.50   &-     &-                     &-     &-      \\ \hline
\end{tabular}
\end{table}
\begin{table}
\caption{Multiplicity in the radiative cooling sample. The K-S test 
result between cumulative distribution functions at different redshifts 
for $\bar{M}_{eff}$ (top) and $M_{max}$ (bottom).}
\begin{tabular}{|c|c|c|c|c|c|}   \hline
{$z$}  &\multicolumn{2}{c|} {Matter} 

       &\multicolumn{2}{c|}{X-ray}   \\ \cline{2-5}

                   	   &{d}
           		   &{p}     
 			   
                   	   &{d}
           		   &{p}	     \\ \cline{1-5} 		
		
0.00 - 0.10   &-     &-                     &-     &-      \\ \cline{1-5}
0.00 - 0.25   &0.17  &4.9 $\times 10^{-2}$  &0.15  &9.6 $\times 10^{-2}$ 
\\ \cline{1-5}		
0.00 - 0.50   &0.26  &3.6 $\times 10^{-4}$  &0.16  &6.9 $\times 10^{-2}$ 
\\ \cline{1-5}		
0.10 - 0.25   &-     &-                     &-     &-      \\ \cline{1-5}
0.10 - 0.50   &0.20  &1.0 $\times 10^{-2}$  &-     &-      \\ \cline{1-5}
0.25 - 0.50   &-     &-                     &-     &-      \\ \hline \hline

{$z$}  &\multicolumn{2}{c|} {Matter} 

       &\multicolumn{2}{c|}{X-ray}   \\ \cline{2-5}

                   	   &{d}
           		   &{p}     
 			   
                   	   &{d}
           		   &{p}	    \\ \cline{1-5} 		
0.00 - 0.10   &-     &-                     &-     &-     \\ \cline{1-5}
0.00 - 0.25   &0.20  &1.0 $\times 10^{-2}$  &-     &-     \\ \cline{1-5}
0.00 - 0.50   &0.27  &2.1 $\times 10^{-4}$  &-     &-     \\ \cline{1-5}
0.10 - 0.25   &-     &-                     &-     &-     \\ \cline{1-5}
0.10 - 0.50   &0.22  &4.4 $\times 10^{-3}$  &-     &-     \\ \cline{1-5}
0.25 - 0.50   &-     &-                     &-     &-     \\ \hline
\end{tabular}
\end{table}
\begin{table}
\caption{Ellipticity in the adiabatic sample. The K-S test result 
between cumulative distribution functions at different redshifts for 
$\epsilon_{eff}$ (top) and $\epsilon_{agg}$ (bottom).}
\begin{tabular}{|c|c|c|c|c|c|}   \hline
{$z$}  &\multicolumn{2}{c|} {Matter} 

       &\multicolumn{2}{c|}{X-ray}   \\ \cline{2-5}

                   	   &{d}
           		   &{p}     
 			   
                   	   &{d}
           		   &{p}	     \\ \cline{1-5} 		
		
0.00 - 0.10   &-     &-                     &-     &-      \\ \cline{1-5}
0.00 - 0.25   &0.16  &5.8 $\times 10^{-2}$  &0.18  &2.9 $\times 10^{-2}$ 
\\ \cline{1-5}		
0.00 - 0.50   &0.16  &8.1 $\times 10^{-2}$  &-     &-      \\ \cline{1-5}
0.10 - 0.25   &-     &-                     &-     &-      \\ \cline{1-5}
0.10 - 0.50   &-     &-                     &-     &-      \\ \cline{1-5}
0.25 - 0.50   &-     &-                     &-     &-      \\ \hline \hline

{$z$}  &\multicolumn{2}{c|} {Matter} 

       &\multicolumn{2}{c|}{X-ray}   \\ \cline{2-5}

                   	   &{d}
           		   &{p}     
 			   
                   	   &{d}
           		   &{p}	    \\ \cline{1-5} 		
0.00 - 0.10   &-     &-                     &0.18  &2.9 $\times 10^{-2}$ 
\\ \cline{1-5}
0.00 - 0.25   &0.16  &5.8 $\times 10^{-2}$  &0.29  &3.5 $\times 10^{-5}$ 
\\ \cline{1-5}		
0.00 - 0.50   &0.22  &2.4 $\times 10^{-3}$  &0.29  &1.9 $\times 10^{-5}$ 
\\ \cline{1-5}		
0.10 - 0.25   &-     &-                     &0.19  &2.0 $\times 10^{-2}$ 
\\ \cline{1-5}
0.10 - 0.50   &0.16  &5.8 $\times 10^{-2}$  &0.18  &2.9 $\times 10^{-2}$ 
\\ \cline{1-5}		
0.25 - 0.50   &-     &-                     &-     &-      \\ \hline
\end{tabular}
\end{table}
\begin{table}
\caption{Ellipticity in the radiative cooling sample. The K-S test 
result between cumulative distribution functions at different 
redshifts for $\epsilon_{eff}$ (top) and $\epsilon_{agg}$ (bottom).}
\begin{tabular}{|c|c|c|c|c|c|}   \hline
{$z$}  &\multicolumn{2}{c|} {Matter} 

       &\multicolumn{2}{c|}{X-ray}   \\ \cline{2-5}

                   	   &{d}
           		   &{p}     
 			   
                   	   &{d}
           		   &{p}	     \\ \cline{1-5} 		
		
0.00 - 0.10   &-     &-     &-     &-                     \\ \cline{1-5}
0.00 - 0.25   &-     &-     &0.16  &6.9 $\times 10^{-2}$  \\ \cline{1-5}
0.00 - 0.50   &-     &-     &0.15  &9.6 $\times 10^{-2}$  \\ \cline{1-5}
0.10 - 0.25   &-     &-     &-     &-                     \\ \cline{1-5}
0.10 - 0.50   &-     &-     &-     &-                     \\ \cline{1-5}
0.25 - 0.50   &-     &-     &-     &-                     \\ \hline \hline

{$z$}  &\multicolumn{2}{c|} {Matter} 

       &\multicolumn{2}{c|}{X-ray}   \\ \cline{2-5}

                   	   &{d}
           		   &{p}     
 			   
                   	   &{d}
           		   &{p}	    \\ \cline{1-5} 		
0.00 - 0.10   &-     &-                     &-     &-      \\ \cline{1-5}
0.00 - 0.25   &-     &-                     &-     &-      \\ \cline{1-5}
0.00 - 0.50   &0.24  &1.7 $\times 10^{-3}$  &0.25  &6.2 $\times 10^{-4}$ 
\\ \cline{1-5}		
0.10 - 0.25   &-     &-                     &-     &-      \\ \cline{1-5}
0.10 - 0.50   &0.22  &4.4 $\times 10^{-3}$  &0.20  &1.0 $\times 10^{-2}$ 
\\ \cline{1-5}		
0.25 - 0.50   &-     &-                     &-     &-      \\ \hline
\end{tabular}
\end{table}
\begin{table}
\caption{The rate of evolution and the error in its measurement 
for clusters in the adiabatic sample. The rate is given by the 
slope of the line obtained from the least-square fit to $M$ vs. $z$ 
and $\epsilon$ vs. $z$ relationship (Fig. \ref{adiabatic_evol}). 
For each measure the 1st and 2nd row show the results obtained, 
respectively, from the median and the mean. The second column for 
each type of clusters shows the error estimated for the slope.} 
\begin{tabular}{|c|c|c|c|c|}   \hline
\label{adiabatic_evol_table}

{Parameter}  &\multicolumn{2}{c|} {Matter} 

       &\multicolumn{2}{c|}{X-ray}   \\ \cline{2-5}

                   	   &{slope}
           		   &{error}     
 			   
                   	   &{slope}
           		   &{error}   \\ \cline{1-5} 

&0.19  &0.07    &0.22  &0.14 \\ \cline{2-5}
\raisebox{+1.5ex}[0pt]{$\bar{M}_{eff}$} &0.22  &0.06 &0.21  &0.09 \\ \hline

&0.50  &0.13    &0.53  &0.27 \\ \cline{2-5}
\raisebox{+1.5ex}[0pt]{$M_{max}$} &0.50  &0.15 &0.58  &0.30  \\ \hline 

&0.02  &0.02    &0.06  &0.04 \\ \cline{2-5} 
\raisebox{+1.5ex}[0pt]{$\bar{\epsilon}_{eff}$} &0.03  &0.02  &0.05  
&0.03 \\ \hline   

&0.18  &0.04    &0.31  &0.07 \\ \cline{2-5}    
\raisebox{+1.5ex}[0pt]{$\bar{\epsilon}_{agg}$} &0.16  &0.03  &0.20  
&0.06 \\ \hline    
\end{tabular}
\end{table}

\begin{table}
\caption{The rate of evolution and the error in its measurement 
for clusters in the radiative cooling sample. The rate is obtained 
from $M$ vs. $z$ and $\epsilon$ vs. $z$ relationship (Fig. 
\ref{radiative_evol}) The presentation style is similar to table 
\ref{adiabatic_evol_table}.}
\begin{tabular}{|c|c|c|c|c|}   \hline
\label{radiative_evol_table}

{Parameter}  &\multicolumn{2}{c|} {Matter} 

       &\multicolumn{2}{c|}{X-ray}   \\ \cline{2-5}

                   	   &{slope}
           		   &{error}     
 			   
                   	   &{slope}
           		   &{error}   \\ \cline{1-5} 

&0.19  &0.02    &0.13  &0.04 \\ \cline{2-5}
\raisebox{+1.5ex}[0pt]{$\bar{M}_{eff}$} &0.32  &0.04 &0.19  &0.04 \\ \hline 

&0.54  &0.03    &0.12  &0.30 \\ \cline{2-5}    
\raisebox{+1.5ex}[0pt]{$M_{max}$} &0.74  &0.02 &0.29  &0.14 \\ \hline 

&0.03  &0.04    &0.02  &0.02 \\ \cline{2-5}     
\raisebox{+1.5ex}[0pt]{$\bar{\epsilon}_{eff}$} &0.01  &0.03 &0.03  
&0.01 \\ \hline   

&0.16  &0.03    &0.22  &0.02 \\ \cline{2-5}    
\raisebox{+1.5ex}[0pt]{$\bar{\epsilon}_{agg}$} &0.14  &0.02 &0.17  
&0.01 \\ \hline 
\end{tabular}
\end{table}

\end{document}